\documentclass[aps,prl,reprint,groupedaddress,showkeys,showpacs]{revtex4-1}
\usepackage{graphicx}
\usepackage{epstopdf}
\usepackage{lineno}
\usepackage[dvipsnames]{xcolor}
\usepackage{amssymb}                    

\newcommand{\bB}{\mathbf{B}}
\newcommand{\bv}{\mathbf{v}}

\newcommand{\bzpm}{\mathbf{z^\pm}}

\begin{document}

\title{Turbulence-driven ion beams in the magnetospheric Kelvin-Helmholtz instability}

\author{Luca Sorriso-Valvo}
\affiliation{Nanotec/CNR, U.O.S. di Cosenza, Ponte P. Bucci, cubo 31C, 87036 Rende, Italy}
\affiliation{Departamento de F\'isica, Escuela Polit\'ecnica Nacional, Quito, Ecuador}
\author{Filomena Catapano}
\affiliation{Dipartimento di Fisica, Universit\`a della Calabria, Ponte P. Bucci, cubo 31C, 87036 Rende, Italy}
\affiliation{LPP-CNRS/Ecole Polytechnique/Sorbonne Universit\'e, Paris, France}
\author{Alessandro Retin\`o}
\affiliation{LPP-CNRS/Ecole Polytechnique/Sorbonne Universit\'e, Paris, France}
\author{Olivier Le Contel}
\affiliation{LPP-CNRS/Ecole Polytechnique/Sorbonne Universit\'e, Paris, France}
\author{Denise Perrone}
\affiliation{Department of Physics, Imperial College of London, London SW7 2AZ, United Kingdom}
\author{Owen W. Roberts}
\affiliation{Space Research Institute, Austrian Academy of Sciences, Schmiedlstrasse 6, 8042 Graz, Austria}
\author{Jesse T. Coburn}
\affiliation{Dipartimento di Fisica, Universit\`a della Calabria, Ponte P. Bucci, cubo 31C, 87036 Rende, Italy}
\author{Vincenzo Panebianco}
\affiliation{Dipartimento di Fisica, Universit\`a della Calabria, Ponte P. Bucci, cubo 31C, 87036 Rende, Italy}
\author{Oreste Pezzi}
\affiliation{Gran Sasso Science Institute, Viale F. Crispi 7, 67100 L’Aquila, Italy}
\affiliation{Dipartimento di Fisica, Universit\`a della Calabria, Ponte P. Bucci, cubo 31C, 87036 Rende, Italy}
\author{Francesco Valentini}
\affiliation{Dipartimento di Fisica, Universit\`a della Calabria, Ponte P. Bucci, cubo 31C, 87036 Rende, Italy}
\author{Silvia Perri}
\affiliation{Dipartimento di Fisica, Universit\`a della Calabria, Ponte P. Bucci, cubo 31C, 87036 Rende, Italy}
\author{Antonella Greco}
\affiliation{Dipartimento di Fisica, Universit\`a della Calabria, Ponte P. Bucci, cubo 31C, 87036 Rende, Italy}
\author{Francesco Malara}
\affiliation{Dipartimento di Fisica, Universit\`a della Calabria, Ponte P. Bucci, cubo 31C, 87036 Rende, Italy}
\author{Vincenzo Carbone}
\affiliation{Dipartimento di Fisica, Universit\`a della Calabria, Ponte P. Bucci, cubo 31C, 87036 Rende, Italy}
\author{Pierluigi Veltri}
\affiliation{Dipartimento di Fisica, Universit\`a della Calabria, Ponte P. Bucci, cubo 31C, 87036 Rende, Italy}
\author{Federico Fraternale}
\affiliation{Dipartimento di Scienza Applicata e Tecnologia, Politecnico di Torino, Torino, Italy}
\author{Francesca Di Mare}
\affiliation{Department of Physics, University of Oslo, Sem S{\ae}lands Vei 26, Blindern, 0316 Oslo, Norway}
\author{Raffaele Marino}
\affiliation{Laboratoire de M\'ecanique des Fluides et d'Acoustique, CNRS, \'Ecole Centrale de Lyon, Universit\'e Claude Bernard Lyon~1, INSA de Lyon, F-69134 \'Ecully, France}
\author{Barbara Giles}
\affiliation{NASA, Goddard Space Flight Center, Greenbelt MD 20771, USA}
\author{Thomas E. Moore}
\affiliation{NASA, Goddard Space Flight Center, Greenbelt MD 20771, USA}
\author{Christopher T. Russell}
\affiliation{Institute of Geophysics and Planetary Physics, and Department of Earth, Planetary, and Space Sciences, University of California, Los Angeles, California, USA}
\author{Roy B. Torbert}
\affiliation{Space Science Center, University of New Hampshire, Durham, New Hampshire, USA}
\author{Jim L. Burch}
\affiliation{Southwest Research Institute, San Antonio, Texas, USA}
\author{Yuri V. Khotyaintsev}
\affiliation{Swedish Institute of Space Physics, Uppsala, Sweden}

\date{\today}

\begin{abstract}
The description of the local turbulent energy transfer, and the high-resolution ion distributions measured by the Magnetospheric Multiscale mission, together provide a formidable tool to explore the cross-scale connection between the fluid-scale energy cascade and plasma processes at sub-ion scales. When the small-scale energy transfer is dominated by Alfv\'enic, correlated velocity and magnetic field fluctuations, beams of accelerated particles are more likely observed. Here, for the first time we report observations suggesting the nonlinear wave-particle interaction as one possible mechanism for the energy dissipation in space plasmas.
\end{abstract}

\pacs{94.05.-a, 94.05.Lk, 95.30.Qd}

\keywords{magnetosphere, turbulence, dissipation}

\maketitle

Space plasmas often provide vivid examples of turbulent, weakly collisional magnetized flows~\citep{living}. 
Among other astrophysical plasmas, those near Earth are particularly important because they can be probed by satellites, which allow for unique {\it in-situ} measurements of electromagnetic fields and particle velocity distribution functions (VDFs). 
Such measurements expose the strongly turbulent nature of the solar wind (SW) and of the terrestrial magnetospheric plasma~\citep{marschtu}. At scales large enough, space plasmas can be described in the fluid magnetohydrodynamic (MHD) approximation~\citep{biskamp}. 
A Kolmogorov-like phenomenology~\citep{K41,frisch} provides predictions for 
anisotropic power-law spectra of magnetic and velocity fluctuations~\citep{GS95}, and intermittency~\citep{frisch}, both broadly supported by observations~\citep{living,marschtu,horbury1997,sorrisovalvo99}. 
The intermittency of the turbulent cascade implies the formation of small-scale structures, such as current sheets, tangential or rotational discontinuities, and vorticity filaments~\citep{brunoLIM,grecoperri,vorticity,perrone16,perrone17}. 
This is the result of inhomogeneous energy transfer, providing a more efficient dissipation of the turbulent energy~\citep{frisch}.
The SW exhibits non-Gaussian statistics at large scales as well, possibly associated with the creation of shears acting as triggers for the onset of turbulent cascades in the interplanetary plasma~\citep{marino_12,nota1}. 

At scales smaller than the proton gyro-radius or inertial length, MHD 
approximations fail, and kinetic processes involving field-particle interactions must be considered. Furthermore, near 1 AU non-Maxwellian VDFs of ions and electrons are measured as expected from the low collision rate of the SW~\citep{marschtu}. 
However, the cross-scale interconnection between processes occurring in the two ranges of scales is still poorly understood~\citep{yang,howes,hermitePRL,hermitePOP}. 
There is growing evidence that the kinetic processes are enhanced in the proximity of the turbulence-generated structures, which carry a larger amount of energy than the surrounding background. For example, ions~\citep{osman,tessein} and electrons~\citep{chasapis2015,chasapis2018a,chasapis2018} are energized in the proximity of the most intense small-scale current sheets. This has also been confirmed in Vlasov-Maxwell numerical simulations~\citep{servidio2012,servidio2014}. 
The processes responsible for the different forms of energization may involve magnetic reconnection~\citep{retino2007,phan2018}, plasma instabilities~\citep{matteini2013,breuillard2016} and enhancement of collisions~\citep{oreste1,oreste2}, and their triggers are a current topic of interest in the community~\citep{chen}.

Investigating turbulent plasma cross-scale processes in depth requires the identification of magnetic and velocity structures in the flow. 
Complementary to the standard techniques, such as the local intermittency measure~\citep{farge,veltri,brunoLIM,perrone17} or the partial variance of increments~\citep{greco2009,grecoperri}, a different heuristic proxy~\citep{marschtu}, related to the local turbulent energy transfer rate across scales, was recently used to identify regions of small-scale accumulation of energy~\citep{sorriso2018a,sorriso2018b}. 
In the MHD approximation, the fluctuations obey the Politano-Pouquet law~\citep{pp98}, which prescribes a linear scaling relation between the 
third-order energy transfer rate and the mean energy dissipation rate, upon homogeneity, scale separation, isotropy, and time-stationarity.
For a plasma time series, using the Taylor hypothesis $r=t\langle v\rangle$ to interchange space ($r$) and time ($t$) arguments via the bulk speed 
$\langle v\rangle$~\citep{taylor}, the basic version of the Politano-Pouquet law for the 
mixed third-order moments $Y^\pm({\Delta t})$ is   
\begin{equation}
Y^\pm({\Delta t}) = \langle |\Delta \bzpm(t,{\Delta t})|^2  \, \Delta z^\mp_{l}(t,{\Delta t}) \rangle = -\frac{4}{3}  \langle \varepsilon^\pm \rangle \Delta t \langle v \rangle \; .
\label{yaglom}
\end{equation}
$\Delta\psi(t,\Delta t)=\psi(t+\Delta t)-\psi(t)$ indicates the increment of a generic field $\psi$ across a temporal scale $\Delta t$, and the subscript $l$ indicates the longitudinal component, {\it i.e.} parallel to the bulk speed; $\bzpm=\bv\pm\bB/\sqrt{4\pi\rho}$ are the Elsasser variables that couple the plasma velocity $\bv$ and the magnetic field $\bB$ expressed in velocity units through the mass density $\rho$. When considering the total energy flux $Y=(Y^++Y^-)/2$, the proportionality factor of the Politano-Pouquet law is the mean energy transfer rate $\langle\varepsilon\rangle=(\langle\varepsilon^+\rangle+\langle\varepsilon^-\rangle)/2$.  
The Politano-Pouquet law has been validated in numerical simulations~\citep{sorriso2002,andres}, in the SW~\citep{mcbride2005,prl,apjl,mcbride2008,marino_11}, where results are compatible with the energy flux necessary to justify the observed plasma heating~\citep{apjl,smith,prlcomp,coburn,marino_11,supratik}, and in the terrestrial magnetosheath~\citep{hadid,bandy,bandy2}.

Based on the law~(\ref{yaglom}), a heuristic proxy of the local energy transfer rate (LET) at the scale ${\Delta t}$ is thus defined by introducing the quantity: 
\begin{equation}
\varepsilon^\pm(t,{\Delta t}) = -\frac{|\Delta  z^\pm(t,\Delta t)|^2\, \Delta z^\mp_{l}(t,\Delta t)}{{\Delta t} \langle v \rangle} \, ,
\label{pseudoenergy}
\end{equation} 
and then computing the average $\varepsilon(t,{\Delta t}) = (\varepsilon^+(t,{\Delta t} ) + \varepsilon^- (t,{\Delta t}))/2$. 
At each scale, the field increments in the time series can thus be associated with the local value of $\varepsilon(t,{\Delta t})$~\citep{marschtu,sorriso2015,sorriso2018a}, assuming smoothness of the fields.  
Moreover, when written in terms of velocity and magnetic field, the LET can be separated in two additive terms, one associated with the magnetic and kinetic energy advected by the velocity fluctuations, $\varepsilon_e=-3/(4\Delta t \langle v \rangle) [\Delta v_{l}(\Delta v^2+\Delta b^2)]$, and the other with the cross-helicity coupled to the longitudinal magnetic fluctuations, $\varepsilon_c=-3/(4\Delta t \langle v \rangle)[-2\Delta b_{l}(\Delta \mathbf{v}\cdot\Delta \mathbf{b})]$~\citep{sorriso2018a,nota2}.

Despite its approximated nature, conditional analysis of temperature profiles in the proximity of LET peaks performed on Helios 2 SW data~\citep{sorriso2018a} and on hybrid Vlasov-Maxwell or fully kinetic particle-in-cell numerical simulations~\citep{sorriso2018b,yang2018} has recently shown that the proxy correctly identifies regions of enhanced kinetic processes, mostly in agreement with standard methods.

In this letter, we use measurements provided by the Magnetospheric Multiscale (MMS) mission~\citep{burch16}. The unprecedented high-cadence for ions~\citep{Pollock16} and magnetic fields~\citep{Russell2016} allows us to explore in depth the link between the MHD energy cascade and the kinetic processes associated with deviations from Maxwellian distribution functions.

On 8 September 2015, MMS was located in the dusk-side magnetopause, moving from the low-latitude boundary layer into the magnetosheath, between 10:07:04 UT and 11:25:34 UT. During this period the spacecraft orbit experienced multiple crossings of the large-scale vortices generated by the Kelvin-Helmholtz (KH) instability. 
Crossings were revealed by several ion-scale periodic current sheets~\citep{Eriksson16}, separating the hotter plasma inside the magnetosphere from the denser boundary layer. 
Turbulence in the boundary layer intervals was studied in depth, showing the presence of a well defined inertial range and intermittency~\cite{Stawarz2016}, after validating the Taylor hypothesis. In this work, we have selected 53 of these boundary layer subintervals, carefully excluding the periodic current sheets and magnetosheath regions based on high temperature and low density, and having relatively stationary fields. This resulted in intervals between 10 s and 150 s long, which provide a non-continuous ensemble of turbulent plasma~\citep{Rossi,Stawarz2016}, with typical ion-cyclotron 
frequency $f_{ci}\simeq 1$ Hz and magnetic fluctuation level $\delta B_{rms}/B_0\simeq 0.15$. The ion plasma $\beta_i=2v^2_{th}/v^2_A$, with the thermal speed $v_{th}=\sqrt{k_B T_p / m_p}$ and the Alfv\'en speed $v_A=B/\sqrt{4\pi\rho}$, is around unity, fluctuating in the range 0.5---1.5.
Magnetic fluctuations display a robust $-5/3$ power-law spectrum in the MHD range of scales (see the supplemental material~\citep{supp1}), approximately between 0.04 and 0.4 Hz, followed by a steeper $-3.2$ spectral exponent in the ion range~\citep{Stawarz2016}. 
Structure function analysis (not shown) reveals that intermittency is also observed.
Substantial electrostatic wave activity was also identified throughout the interval~\citep{Stawarz2016,electrostatic}.

The proxy $\varepsilon(t,\Delta t)$ given in Eq.~(\ref{pseudoenergy}) was computed at different scales $\Delta t$ using the MMS1~\citep{mms1} spacecraft velocity, magnetic field and density measurements, for the turbulent regions of the 53 sub-intervals described above~\citep{Stawarz2016}. 
Note that the sample under analysis is generally compressible. Based on recent results, compressibility should result in enhanced transfer in the locations where compressive effects are stronger~\cite{prlcomp,andres,hadid}. Nevertheless, here we use the incompressible proxy as a first-approach approximation, deferring the extension to a more complete, compressible version to future work. 
Measurements of the ion distribution functions and moments are provided by the Fast Plasma Investigation (FPI) instrument~\citep{Pollock16}, covering an energy range of [0.1---30] keV, with cadence of 150 ms. Magnetic field were measured by the Flux-Gate Magnetometers (FGM)~\citep{Russell2016}, with a cadence of 128 Hz, and were carefully synchronized to the plasma data.
The local longitudinal direction was determined as the average speed evaluated over 30 s running windows, of the order of the velocity correlation scale~\citep{Stawarz2016}. 
In the following, we will focus on the scale $\Delta t=1.2$~s, located near the transition between the fluid and the ion kinetic scales~\citep{Stawarz2016}. 
At such scales, the third-order law is still valid, so that the local proxy LET gives a reasonable description of the rate at which energy is locally transferred, being available to excite smaller scales processes. Note that the LET is indicative of non-linear transport and does not include the possible eddies temporal distortion.
In order to simplify the notation, the LET explicit $t$ and $\Delta t$ dependency will be dropped in the following.

\begin{figure}
  \begin{center}
  \includegraphics[width=\columnwidth]{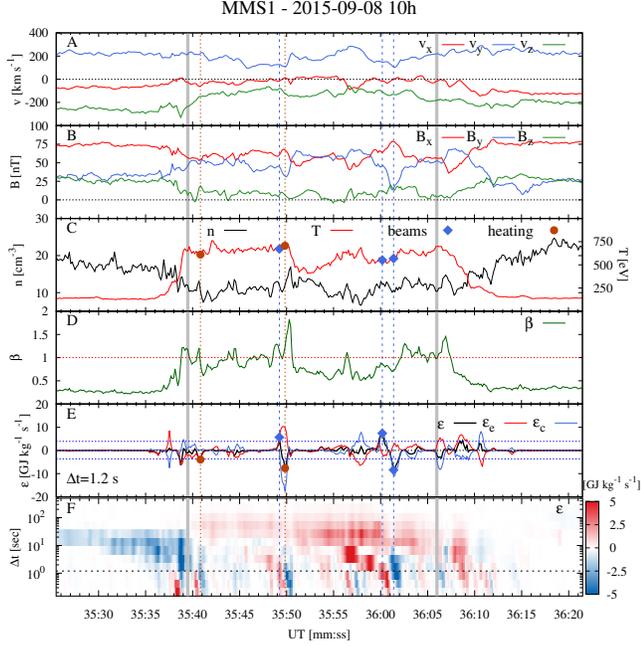}
  \caption{A one-minute subsample from the MMS1 data interval, starting at 10:35:21.359 UT on 2015-09-08. Thick vertical gray lines enclose one of the subintervals used for the analysis. Panel A: velocity components (Geocentric Solar Magnetospheric frame, GSM); B: magnetic field components (GSM); C: ion density and temperature; D: ion plasma $\beta_i$; E: $\varepsilon$, $\varepsilon_e$ and $\varepsilon_c$ at $\Delta t=1.2$ s, with the indication of the two thresholds $\theta^+\sigma$ and $\theta^-\sigma$ as blue horizontal dotted lines; F: the scalogram of $\varepsilon$, the horizontal dashed lines indicating the scale $\Delta t=1.2$ s. The dashed or dotted vertical lines in all panels and the markers in panels C and E indicate the VDFs observed for this subinterval, separately for beams (blue diamonds and dashed line) and heating (dark-orange circles and dotted line).}
  \label{mmsdata}
  \end{center}
  \end{figure}

Panels (A)--(D) of Figure~\ref{mmsdata} show MMS measurements of several quantities in one of the 53 selected BL subintervals. 
Panel (E) illustrates the bursty, intermittent nature of $\varepsilon$. A representation of the energy flow across scales is provided by the scalogram of the LET, shown in panel (F). 
The energy path across scales is clearly visible, as well as the small-scale intermittent structures (the bright regions at small scales) that contain a large fraction of energy. 
Intense, small-scale LET events often present a double channel of positive-negative energy flux (see {\it e.g.} around t=36:01), revealing the complexity of the energy transport mechanism~\citep{coburn2014,camporeale2018}. 

%
%
%
Upon averaging over the whole ensemble of 53 sub-intervals, the scale-dependent third-order moment~(\ref{yaglom}) is approximately in agreement with the linear prediction~(\ref{yaglom}), as evidenced in the supplemental material~\citep{supp2}, and provides a mean energy transfer rate $\langle \varepsilon \rangle \simeq 53\pm 8$ MJ kg$^{-1}$s$^{-1}$, compatible with previous observations in the magnetosheath~\citep{hadid}. To our knowledge, this is the first observation of the Politano-Pouquet law inside the Earth magnetospheric boundary layer. 
Notice that the standard deviation of the LET at the bottom of the inertial range ($\Delta t=1.2$ s) is $\sigma=3016$ MJ kg$^{-1}$s$^{-1}$, indicating that the local flux fluctuations are much larger than the average energy flux estimated through equation~(\ref{yaglom}). This suggests an analogy between LET and the highly fluctuating transfer functions obtained from the nonlinear term of the fluid equations, whose integral provides the average energy flux~\cite{alexakis,marino_14}. 

\begin{figure}
  \begin{center}
    \includegraphics[width=\columnwidth]{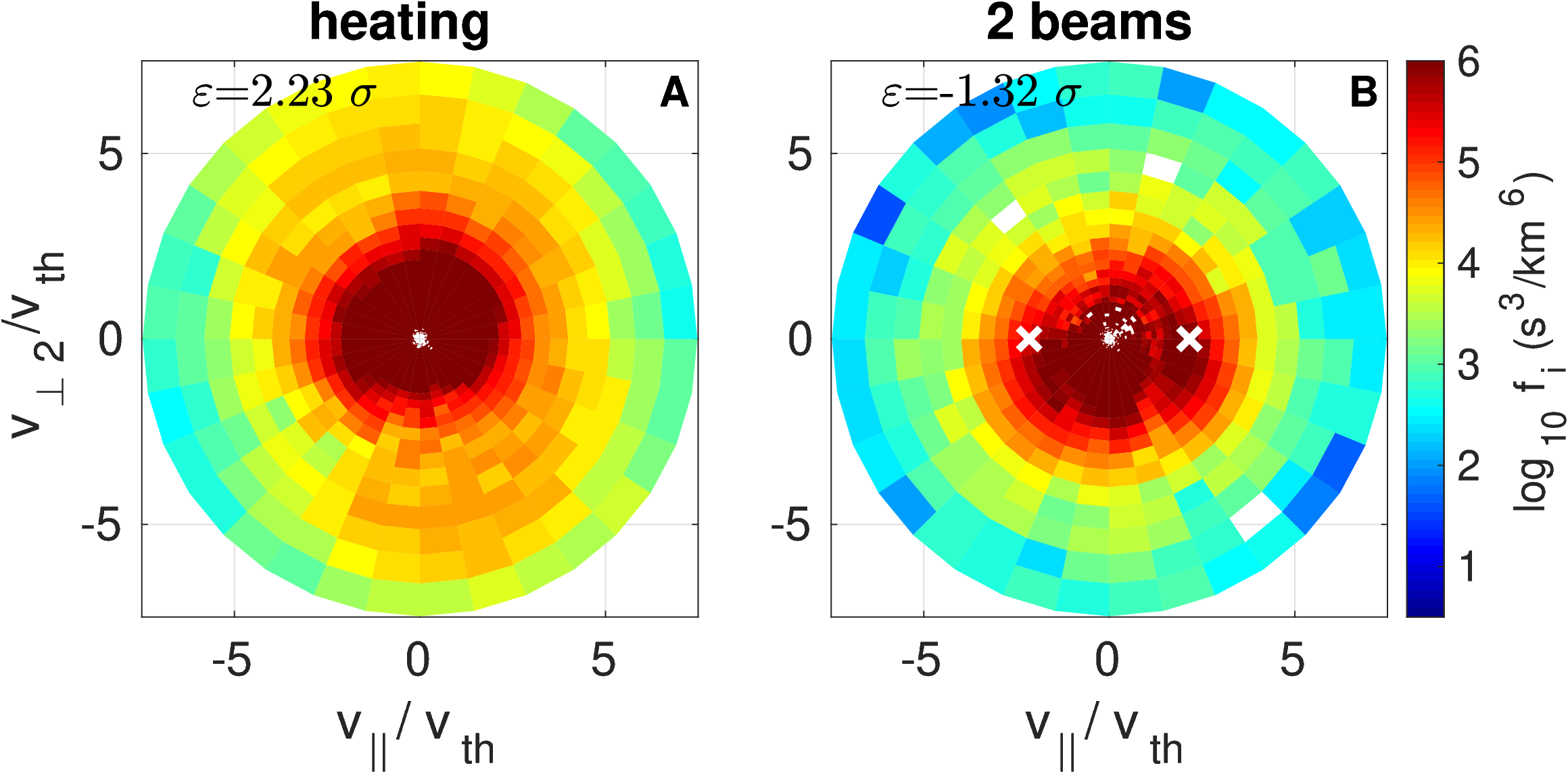}
  \end{center}
  \caption{Examples of 2D cuts of the 3D ion VDF, measured at LET peaks at 10:42:32.78 UT (A) and 10:07:45.82 UT (B). Here $v_{\parallel}$ is directed along the local magnetic field and $\textbf{v}_{\perp,2}=\textbf{\^v}\times(\textbf{\^v}\times\textbf{\^b})$, where $\textbf{\^v}=\textbf{v}/|\textbf{v}|$ and $\textbf{\^b}=\textbf{B}/|\textbf{B}|$.
In each panel, the type of VDF is indicated, along with the LET value in standard deviation units. Axes are normalized to the thermal velocity $v_{th}$. The white crosses in panel (B) represent the local value of the normalized Alfv\'en velocity $v_{A}$.} 
  \label{fig-vdfs}
  \end{figure}

In order to investigate the connection between the turbulent energy being 
transferred towards small scales and the deformation of the ion VDF at smaller scales, and therefore to provide evidence of the feedback of fluid on kinetic dynamics, we identified 94 positive and 94 negative peaks of LET by setting the two thresholds $\varepsilon>\theta^+\sigma$ and $\varepsilon<\theta^-\sigma$. Here $\theta^+=1.3$ and $\theta^-=-1.2$ are the threshold values in units of LET standard deviation, the subscripts indicating the positive or negative LET ensemble. At the time of each peak, the ion VDF was smoothed over 0.45 s (i.e. averaging over three data points) in order to reduce measurement noise, and then normalized to the local thermal speed $v_{th}$. 
Two-dimensional cuts of each VDF were visually examined in order to identify possible features and deviation from Maxwellian. All selected VDFs were then classified according to the following categories: ($i$) quasi-Maxwellian; 
($ii$) presence of broad particle energization (here simply labeled as ``heating''~\citep{nota3}); 
($iii$) presence of one or two beams~\cite{nykyri,li2016,vernisse2016}; 
($iv$) other uncategorized features. 
Examples of classes ($ii$) and ($iii$) are visible in the two-dimensional cuts in the $v_{\perp 2}$-$v_{\parallel}$ plane shown in Figure~\ref{fig-vdfs}, where the velocity components are with respect to the local magnetic field.
one of the events above the threshold presents Maxwellian VDF (see Table~\ref{table1}). 
Broad particle energization (panel A) is the most common feature (more than two-thirds of the cases), while beams (panel B) are clearly visible in about 27\% of the cases. Note that beams are more likely generated by a positive local energy transfer. 

\begin{figure}
  \begin{center}
  \includegraphics[width=0.8\columnwidth]{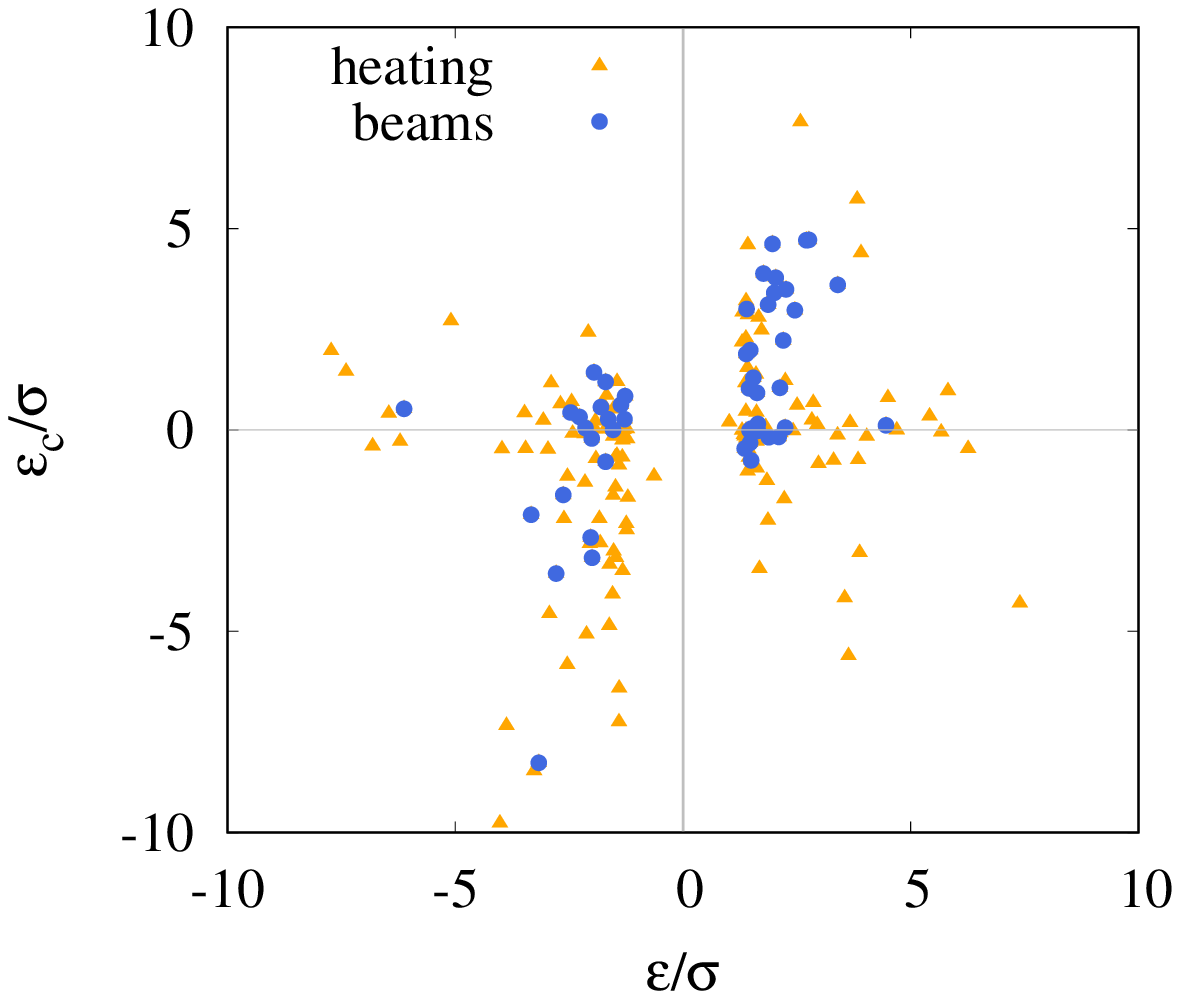}
  \includegraphics[width=0.8\columnwidth]{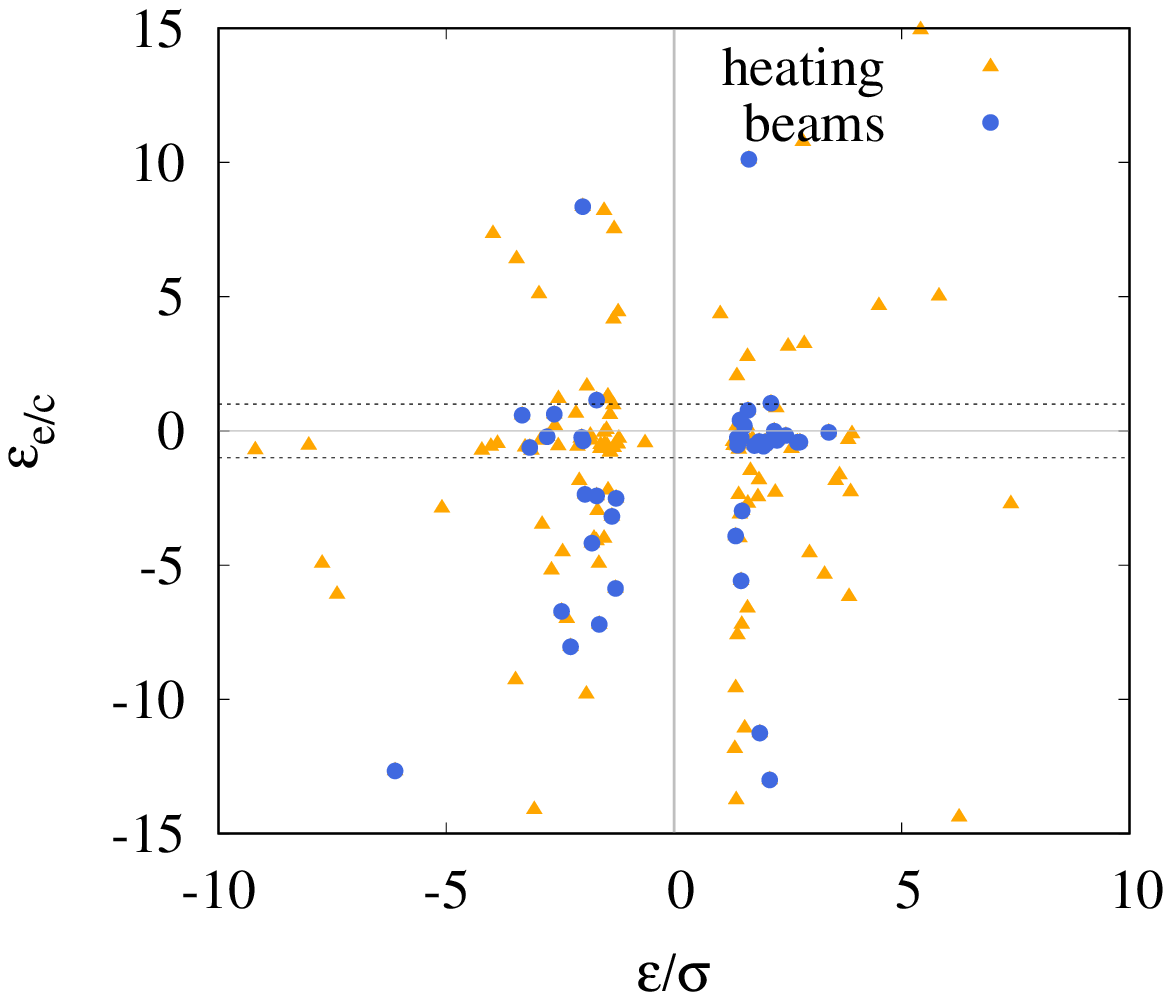}
  \end{center}
  \caption{Distribution of VDFs in the $\varepsilon$--$\varepsilon_c$ (top) and in the $\varepsilon$--$\varepsilon_{e/c}$ (bottom) planes, highlighting the majority of beams for dominating, positive $\varepsilon_c$ (blue circles) as opposed to the more spread heating (orange triangles).}
  \label{fig-beams}
  \end{figure}

In order to compare the statistics with occurrence rates corresponding to small LET values, we have randomly selected 188 VDFs with $|\varepsilon|<10^{-3}\sigma$. 
More than half of these are roughly quasi-Maxwellian, confirming that lower energy transfer results in weaker deviation from Maxwellian; heating is seen for about one fourth of the cases, and only one sixth show presence of beams.  
Results shown in Figure~\ref{fig-vdfs} and collected in Table~\ref{table1} demonstrate that the particle VDFs are characterized by more evident non-Maxwellian features in the proximity of larger turbulent energy transfer~\citep{osman,chasapis2015,chasapis2018a,chasapis2018,vorticity,greco12,valentini2016}. 
%
%
%
%
%
\begin{table} 
\begin{center}
\caption{Occurrence rate of each VDF class measured at positive and negative LET peaks and for $|\varepsilon|<10^{-3}\sigma$.}
\vskip 12pt
\begin{tabular}{c|c|c|c} 
	Classes  &  $ |\varepsilon|\sim 0 $  &  $ \varepsilon > \theta^+\sigma $  &  $ \varepsilon < \theta^-\sigma $       \\ 
	\hline
    \hline
	q-Maxwellian  & 0.57 & 0.00 & 0.00 \\ 
    Heating       & 0.26 & 0.63 & 0.76 \\ 
    Beams         & 0.17 & 0.33 & 0.21 \\ 
    Other         & 0.00 & 0.04 & 0.03 \\ 
   \hline
\end{tabular}
\label{table1}
\end{center}
\end{table}
%
%
Unlike the other aforementioned proxies, the ratio $\varepsilon_{e/c}=\varepsilon_e/\varepsilon_c$ allows to establish whether the cascading energy driving the kinetic processes is dominated by strong gradients, such as current sheets and vorticity filaments ($|\varepsilon_{e/c}|>1$, found in about two thirds of the cases), or rather by Alfv\'enic-like, aligned fluctuations ($|\varepsilon_{e/c}|<1$, as in one third of the cases). Figure~\ref{fig-beams} shows the distribution of VDFs with beams or heating as a function of the total ($\varepsilon$) and partial ($\varepsilon_c$ or $\varepsilon_{e/c}$) energy transfer rates. 
%
%
%

The top panel shows that heating is increasingly dominating for larger energy transfer, while most of the beams are approximately limited to $1\sigma\lesssim|\varepsilon|\lesssim 3\sigma$. This seems to indicate that particularly intense energy transfer may prevent the generation of ordered particle energization, such as beams. A closer look reveals that the large majority of beams are observed for positive cross-helicity contribution $\varepsilon_c > 0$ (overall $\sim 73$\%, including $\sim 80$\% positive and $\sim 60$\% negative LET peaks).

Looking at the ratio between the energy and cross-helicity terms (bottom panel), in the cases with positive energy transfer the beams are predominantly seen for $|\varepsilon_{e/c}|<1$ ({\it i.e.} within the two horizontal dotted lines). 

Therefore, while highly energetic, uncorrelated current and vorticity structures produce mostly disordered particle energization, the generation of beams seems to be mainly associated with the presence of Alfv\'enic velocity and magnetic fluctuations carrying energy towards smaller scales. 

Note that beams were mostly observed to be magnetic-field aligned (92\% of the cases), and robustly located at $v_{beam}\simeq \pm v_A$, the mean ratio being $V_{beam}/V_A=0.98\pm0.09$, where the error is the standard deviation.
Furthermore, for most of the beams (although not exclusively), localized ion-cyclotron wave activity was detected, as left- and right-handed polarized magnetic fluctuations were identified through wavelet phase difference and coherence analysis~\citep{supp3}. The presence of Alfv\'enic vortex-like structures was also observed at the beams~\citep{supp4}. Finally, high-frequency electrostatic activity~\citep{electrostatic} was preliminarily observed in correspondence with several VDFs with beams~\citep{supp5}. 

These observations point to a possible interpretation in terms of beams being generated by resonant interaction of protons with Alfv\'enic-like fluctuations. 
From quasi-linear theory, a diffusive plateau in the longitudinal proton velocity distribution is generated as the result of resonant wave-particle interaction~\cite{kennel66}.
In the nonlinear case, for large amplitude fluctuations, the plateau is replaced by a bump along the magnetic field direction~\cite{valentini08,pezzi2017}, associated with a significant level of electrostatic activity~\cite{valentini11}.  Moreover, if particles interact with fluctuations of the ion-cyclotron branch, the beam is located at $v_\parallel \simeq V_A$~\cite{valentini09,valentini11}. Some of these features were observed in the present MMS data analysis, while similar results were observed for the electron VDFs~\citep{graham}. Note that the interaction of a beam with the plasma background may also produce streaming instabilities~\citep{wentzel74}.
Strikingly similar results were also observed in a preliminary study of high resolution, two-dimensional Hybrid Vlasov-Maxwell numerical simulations~\citep{1024}, as shown in the supplemental material~\citep{supp6}. This supports the scenario of nonlinear wave-particle interaction as one of the possible mechanisms removing energy from the turbulent cascade. 

The cross-scale coupling between fluid turbulence and kinetic processes has been studied though the high-resolution plasma measurements recorded by the MMS spacecraft during an extended observation period of Kelvin-Helmoltz vortices at the Earth magnetopause boundary layer. Inspired by the third-order law, a heuristic proxy has been used to identify regions of large energy transfer in the time series, where the specific features of the ion VDFs have been examined. Despite the many underlying approximations, the simplified descriptor used here is able to successfully localize regions of BL plasma with ion VDFs that have more pronounced non-Maxwellian features, corresponding to larger energy transfer. 
More in particular, field-aligned beams at $V_A$ are more likely generated when such energy is predominantly carried by Alfv\'enic, aligned velocity and magnetic fluctuations, suggesting the possible role of turbulence-driven Landau resonance in the energy dissipation processes. 
The results presented here thus expose the strong connection between the local details of the inertial-range turbulent energy transfer and its transformation through small-scale kinetic processes in non-collisional space plasmas, which is of broad interest for astrophysical plasmas. Additionally, they advance the knowledge of one of the major open questions in space plasma physics, namely what are the mechanisms responsible for the dissipation of turbulent energy.

The simple MHD-scale proxy used here could also be considered as a estimator of likelihood for the localization of VDFs with the presence of parallel beams. Indeed, when both conditions of a positive peak in the local energy transfer rate ($\varepsilon>\theta^+$), and a dominating cross-helicity term ($\varepsilon_{e/c}<1$) are satisfied, then there is a robust 53\% probability of having one or two parallel beams in the ion VDFs.
These results may thus be relevant for current and future space plasma missions such as MMS, Parker Solar Probe and Solar Orbiter, both for the interpretation of the observations, and as a possible trigger for plasma distributions burst mode and telemetry.

The path towards future steps to improve the proposed diagnostics includes: the use of high-resolution Vlasov numerical simulations; the extension of the third-order law to small-scale dynamics (Hall-MHD and Vlasov); the inclusion of compressive and anisotropy effects; the study of turbulence in the open solar wind (as soon as MMS data are available) and in other space plasma systems; the definition of automated, quantitative techniques to determine the VDF type; and the determination of the causality relationship between the observed beams and reconnection sites~\cite{moore2016,moore2017}. 

\begin{acknowledgments}
We are thankful to Emiliya Yordanova for useful discussion. Work by DP was supported by STFC
grant ST/N000692/1. FV, OP and SP were supported by contract ASI-INAF 2015-039-R.O. ``Missione M4 di ESA: Partecipazione Italiana alla fase di assessment della missione THOR''. RM acknowledges support from the program PALSE (Programme Avenir Lyon Saint-Etienne) of the University of Lyon, in the framework of the program {\it Investissements d'Avenir} (ANR-11-IDEX-0007).
U.S. co-authors are supported by the National Aeronautics and Space Administration (NASA) Magnetospheric Multiscale Mission (MMS) in association with NASA contract NNG04EB99C. We thank the entire MMS team and instrument leads for data access and support. The data presented in this paper are the L2 data of MMS and can be accessed from the MMS Science Data Center (https://lasp.colorado.edu/mms/sdc/public/).
\end{acknowledgments}



\begin{thebibliography}{}
\bibitem[\protect\citeauthoryear{Nota 1}{nota1}]{nota1}
Large and small-scale intermittency, emergence of shears and anomalous dissipation are rather universal features of anisotropic turbulent flows in nature, being also observed in geophysical fluids~\citep{marino_13,marino_15,feraco_18}
\bibitem[\protect\citeauthoryear{NOTA4}{nota4}]{mms1}
At the time of the observation, MMS1 position was optimal for the observation of the current sheets crossings~\citep{Eriksson16}, with respect to the other spacecraft.
\bibitem[\protect\citeauthoryear{NOTA3}{nota3}]{nota3}
The word ``heating'' is used here without implication of actual particle temperature increase, which is of difficult definition for non-Maxwellian plasmas, but rather as a label for more general particle energization, in terms of broadening (either isotropic or anisotropic) of the particle distribution function.
\end{thebibliography}
\end{document}